\documentclass{article}

\usepackage{graphicx}
\usepackage{authblk}
\title{AI Techniques for\\ Software Requirements Prioritization}

\author[1]{Alexander Felfernig}
\affil[1]{\small Applied Software Engineering \& AI, Graz University of Technology, Inffeldgasse 16b/2, A-8010 Graz, Austria}

\date{\small Preprint, cite as: A. Felfernig, AI Techniques for Software Requirements Prioritization. In: M. Kalech and R. Abreu and M. Last (Eds), Artificial Intelligence Methods for Software Engineering (pp. 29-47). World Scientific, 2021.}

\begin{document}

\maketitle

\abstract{Aspects such as limited  resources,  frequently changing market  demands,  and  different technical  restrictions  regarding  the implementation of software requirements (features) often demand for the prioritization of requirements. The task of prioritization is the ranking and selection of  requirements  that  should  be  included  in  future  software  releases.  In this context, an intelligent prioritization decision support is extremely important. The prioritization approaches discussed in this paper are based on different Artificial Intelligence (AI) techniques that can help to improve the overall quality of requirements prioritization processes.









\section{Introduction}

Limited resources, market demands, and technical restrictions regarding the implementation of software features often demand for the prioritization of requirements \cite{Achimugu2014,karim2014bi,Lehtola2004,Mobasher2011}. The focus of prioritization is the \emph{ranking and selection of requirements that should be included in future software releases}.  Intelligent decision support in prioritization is extremely important since especially when dealing with large assortments of requirements, manual prioritization processes tend to become very costly \cite{Alenezi2013,Perini2009,ruhe2002software,Xuan2012}. Potential sub-optimal prioritizations can lead to different negative effects such as \emph{waste of time} due to a  focus on irrelevant requirements, \emph{opportunity costs} due to the fact that the relevant features are not provided first, and \emph{missing focus on market demands} that could lead in the worst case to total loss \cite{Firesmith2004}. In this context, prioritization can take place on the strategic level as well as an on the operative level, which is typically associated with short-term prioritization tasks \cite{Ameller2017,88_Ruhe2005}. The prioritization approaches discussed in this paper are based on AI techniques from the areas of constraint reasoning \& optimization \cite{Tsang1993}, utility-based recommendation \cite{felfernigburke08}, content-based recommendation \cite{77_Pazzani1997}, matrix factorization \cite{Koren2009}, conflict detection \cite{52_Junker2004}, and model-based diagnosis \cite{FelfernigSchubertZehentner2012}.

An overview of different prioritization tasks is given in Figure \ref{fig:prioritizationtasktypes}. This categorization is based on two dimensions. First, \emph{level of requirements} specifies the granularity of requirements specifications, i.e., to which extent these requirements can already be translated into corresponding detailed software features.  Second, \emph{inclusion of constraints} refers to which extent relationships between requirements and relationships to external factors are taken into account in the prioritization process. Examples of constraints  (dependencies) between requirements are $x$ \emph{requires} $y$ ($x$ must not be implemented before $y$) and $x$ \emph{excludes} $y$ (only one of these requirements should be implemented). Examples of external factors are the available budget for a software project, available personnel resources, and specific preferences of stakeholders engaged in a software project. Along with these two dimensions, there exist different prioritization approaches, which can be differentiated with regard to the \emph{granularity level of requirements} and the \emph{degree of the inclusion of constraints}. 

\begin{figure}[ht]
\centering
\includegraphics[scale=0.5]{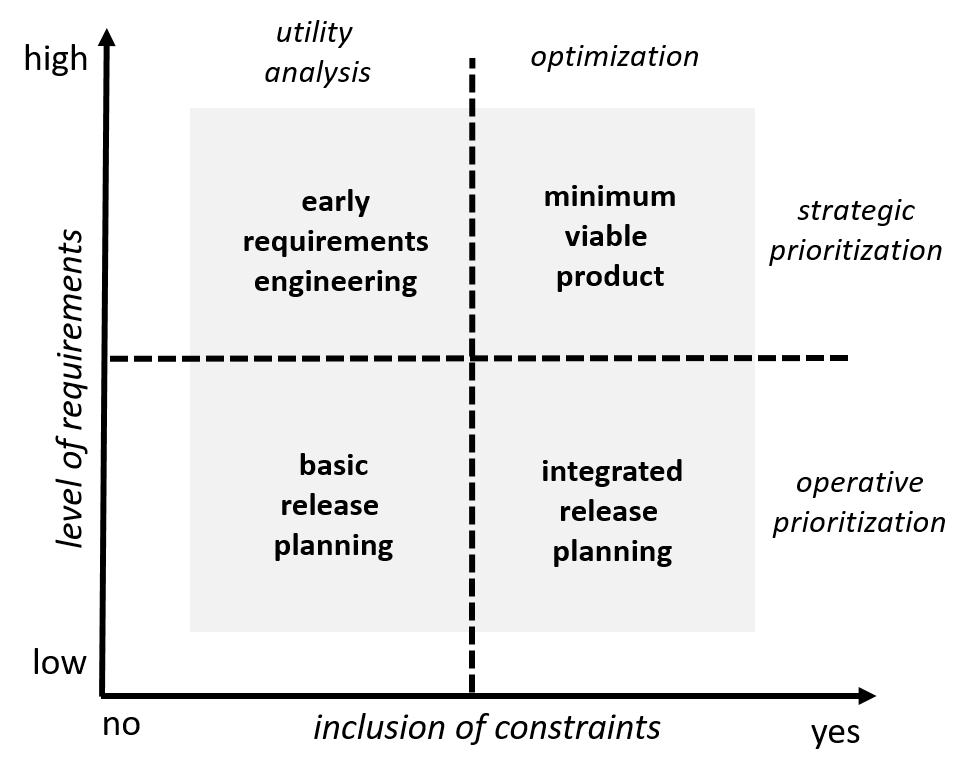}  
\caption{Prioritization variants in software development contexts.} 
\label{fig:prioritizationtasktypes}
\end{figure}

\emph{Early requirements engineering} is related to the idea of figuring out the requirements that have the highest importance, for example, for the market or specific customer communities. Prioritization tasks typically refer to high-level requirements, furthermore, no specific constraints are included. The major focus is to figure out the most relevant features of a product with a market relevance. Requirements in such scenarios can be regarded as high-level, for example, \emph{"the new e-learning software should include a motivation functionality that persuades students to intensively learn the course topics"} or \emph{"the new e-learning software should support natural language based interaction mechanisms"}. 

A \emph{minimal viable product} (MVP) should include a minimal set of features that can be integrated as parts of a fully operable software offered to customers. MVPs are a typical approach to get to the market as soon as possible with the most relevant features of a software. In this context, constraints play an important role since the prioritization of requirements has to take into account constraints such as available personnel and budget resources. Requirements can also be regarded as high-level and constraints primarily refer to available budgets and personnel resources. Examples of constraints are \emph{"motivation features of the new e-learning software should solely include the aspect of social influence"} and \emph{"for the first version of the software, natural language interaction should support the answering of multiple-choice questions with single correct answers"}. 

\emph{Basic release planning} does not fully take into account further constraints such as available budget, personnel resources, and time restrictions regarding the implementation of requirements. This type of prioritization  covers implementation scenarios where releases are planned on an operational level without taking into account in detail constraints regarding available personnel and budget resources as well as time limitations. Examples of requirements in such contexts are \emph{"the basic scenario for a social influence based persuasion is the following ... the user interface implementation of this function should look like as follows ..."} or \emph{"the basic scenario for supporting multiple choice questions in the context of natural language interactions is the following ... the user interface implementation can be sketched as follows ..."}. On a technical level, basic release planning can be performed using approaches similar to those used in the context of early requirements engineering.

Finally, \emph{integrated release planning} represents a full-fledged release planning \cite{nayebi2015analytical,ruhe2010product} on the basis of detailed constraints representing organizational data and rules. In this context, both, constraints regarding dependencies between requirements as well as constraints related to external factors are taken into account. Similar to basic release planning, requirements are defined with fine granularity. A major difference between basic release planning and integrated release planning is the availability of more detailed constraint information, for example, integrated release planning is able to take into account the individual availability of developers (in terms of engagement in other projects and presence or absence during specific time periods). Furthermore, dependencies between requirements can be taken into account on a formal level. 

On the level of prioritization techniques, there are \emph{two basic approaches} to support prioritization processes -- see Achimugu et al. \cite{Achimugu2014}. \emph{First},  prioritization can be regarded as an \emph{optimization task} where the objective is to identify a prioritization that takes into account the preferences of individual stakeholders and also helps to optimize the prioritization with regard to a set of predefined constraints \cite{Kifetew2017}. On a technical level, optimization-based prioritization is often based on a hybrid approach where the identification and aggregation of stakeholder preferences is supported by utility analysis \cite{Adomavicius2010,Huang2011,Wiegers2003} and optimization is performed on the basis of constraint reasoning \cite{Felfernigetal2014,Tsang1993}.

\emph{Utility-based approaches}  focus  on an analysis of the given requirements with regard to a set of \emph{interest dimensions} and less on automated optimization. Different variants of this approach can be implemented, for example, a utility-based ranking can be extended with the concepts of liquid democracy \cite{Atas2018liquid}. Finally, social networks can be exploited as data sources for the identification of new requirements, which are regarded as relevant by the underlying social network \cite{Williams2017}. In terms of the application of the mentioned prioritization techniques, early requirements engineering and basic release planning focus more on utility-based prioritization approaches whereas minimum viable product and integrated release planning focus on optimization-based prioritization approaches. 

The major contributions of this paper are the following. First, we provide an overview of existing techniques that help to improve the quality of prioritization processes in requirements engineering. Second, we show the application of these techniques in  the context of working examples. Third, in order to stimulate further work in related fields, we discuss relevant issues for future work.

The remainder of this paper is organized as follows. Sections \ref{sec:earlyre}--\ref{sec:irp} include a discussion of the application of AI techniques in the scenarios of \emph{early requirements engineering}, \emph{minimum viable products}, \emph{basic release planning}, and  \emph{integrated release planning}. These sections include a description of the underlying scenarios and  working examples. Section \ref{sec:stakeholderrecommendation} provides insights how to support stakeholder selection, which is an important issue when it comes to the assignment of requirement validation tasks. Section \ref{sec:researchissues} provides an overview of issues for future research. Finally, we conclude the paper with Section \ref{sec:conclusions}.

\section{Early Requirements Engineering}\label{sec:earlyre}

A basic means to support prioritization tasks in early requirements engineering is to perform a utility analysis of a given set of requirements. \emph{Utility-based prioritization} is based on the concepts of \emph{multi-attribute utility theory} \cite{Dyer2005} -- different variants thereof are possible. First, individual requirements are  evaluated with regard to \emph{interest dimensions} (e.g., risk level of a requirement and the commercial relevance of a requirement). The utility of the requirement is then determined on the basis of the sum of interest dimension specific utility values. Interest dimensions can be associated with a  weight, for example, \emph{low risks are more important than high profits}. Utility-based prioritization can also be implemented on the basis of analytic hierarchy process (AHP) \cite{Karlsson1997}. A major disadvantage of this approach is that requirements have to be evaluated pairwise which does not scale well when the number of requirements increases.

Interest dimensions, i.e., basic evaluation criteria for utility-based prioritization can differ depending on the underlying decision scenario. Examples of such interest dimensions in company-related software projects are  \emph{effort to implement a requirement}, \emph{risk of not being able to implement a requirement}, and \emph{business relevance of a requirement (profit)}  \cite{Achimugu2014}. In open source settings, the dimensions can be different since  open source contributors have to decide individually on which requirement to work next. Examples of related interest dimensions could be  \emph{personal expertise of an open source developer} and \emph{importance of a requirement for the community} \cite{Felfernig2018RE}. 

Utility analysis  supports stakeholders in the prioritization of requirements with regard to a set of interest dimensions $D = \{d_1, d_2, ..., d_n\}$. The underlying idea is that requirements are first analyzed by individual stakeholders (also denoted as users) -- see Tables \ref{tab:groupbasedmautrelevance}--\ref{tab:groupbasedmautrisk}. Such decisions are often group decisions where  stakeholders are in charge of prioritizing a set of requirements \cite{Felfernig2018}.

In this simplified example, \emph{users} are in charge of evaluating the requirements $req_1 .. req_5$ with regard to the interest dimensions \emph{business relevance} and \emph{risk}. Thereafter, individual evaluations are aggregated to determine the utility of  requirements. In this context, Formula \ref{eq:reqspecificutility} can be used to calculate the utility of a requirement with regard to a specific interest dimension $d$. Furthermore, Formula \ref{eq:overallutility} is used to determine the overall utility of a requirement.

\begin{table}[ht] \begin{center}
\caption{Evaluation of the dimension \emph{relevance}  (high rating = high relevance). \vspace{0.1cm}}
\begin{tabular} {|c|c|c|c|c|}
\hline 
 & user$_{1}$ &  user$_{2}$ & user$_{3}$ &  user$_{4}$\tabularnewline
\hline
\hline 
req$_{1}$ & 1 & 4 & 5 & 2\tabularnewline
\hline 
req$_{2}$ & 10 & 6 & 1 & 7\tabularnewline
\hline 
req$_{3}$ & 2 & 6 & 5 & 2\tabularnewline
\hline 
req$_{4}$ & 1 & 1 & 3 & 7\tabularnewline
\hline 
req$_{5}$ & 7 & 8 & 6 & 5\tabularnewline
\hline 
\end{tabular} \end{center}
\label{tab:groupbasedmautrelevance}
\end{table}

\begin{table}[ht] \begin{center}
\caption{Evaluation of the  dimension \emph{risk} (high rating = low risk). \vspace{0.1cm}} 
{\begin{tabular}{|c|c|c|c|c|}
\hline 
 & user$_{1}$ &  user$_{2}$ & user$_{3}$ &  user$_{4}$\tabularnewline
\hline
\hline 
req$_{1}$ & 2 & 7 & 3 & 2\tabularnewline
\hline 
req$_{2}$ & 9 & 9 & 1 & 7\tabularnewline
\hline 
req$_{3}$ & 2 & 10 & 3 & 2\tabularnewline
\hline 
req$_{4}$ & 2 & 5 & 3 & 1\tabularnewline
\hline 
req$_{5}$ & 3 & 2 & 3 & 5\tabularnewline
\hline 
\end{tabular}} \end{center}
\label{tab:groupbasedmautrisk}
\end{table}

\begin{equation}\label{eq:reqspecificutility}
    utilityreq(req,d)=\frac{\Sigma_{u \in Users} eval(req,d,u)}{|Users|}
\end{equation}

\begin{equation} \label{eq:overallutility}
    utility(req)=\frac{\Sigma_{d \in Dims} utilityreq(req,d) \times weight(d)}{|Dims|}
\end{equation}

 The determined utilities are encoded in a ranking (see Table \ref{tab:groupbasedmautrank}).

\begin{table}[ht] \begin{center}
\caption{ Prioritization of requirements $req_1 .. req_5$ with regard to the interest dimensions \emph{relevance} (weight = 0.75) and \emph{risk} (weight = 0.25). \vspace{0.1cm}}
\begin{tabular}{|c|c|c|c|c|c|}
\hline 
requirement $req_i$ & req$_{1}$ &  req$_{2}$ & req$_{3}$ &  req$_{4}$ &  req$_{5}$\tabularnewline
\hline
\hline 
utility($req_i$) & 4.63 & \bf5.75 & 4.06 & 2.94 & 4.56\tabularnewline
\hline 
priority($req_i$) & 2 & 1 & 4 & 5 & 3\tabularnewline
\hline 
\end{tabular} \end{center}
\label{tab:groupbasedmautrank}
\end{table}

The presented approach to group-based multi attribute utility analysis \cite{Felfernig2018} is based on the assumption that each stakeholder is able to provide feedback on each of the given requirements. This might not be possible for various reasons, for example, stakeholders are simply not available, i.e., do not have time or they might have issues in terms of missing knowledge needed to evaluate a requirement. In such cases, mechanisms are needed to be able to transfer votes in a flexible fashion. Such an approach to liquid-democracy based prioritization is introduced in \cite{Atas2018liquid}. The major difference compared to the aforementioned approach is that individual stakeholders are allowed to vote more than once and to transfer their votes to other stakeholders.

An alternative approach to handle \emph{missing values} in requirements evaluation is to apply machine learning concepts, which help to automatically complete a potentially sparse rating matrix \cite{Koren2009}. The automatically determined requirements evaluations can then be proposed to stakeholders and can also serve as indicators of potential issues related to contradictory evaluations, which have to be resolved. Table \ref{tab:useritemratingtable} depicts a user-item matrix, which includes a couple of missing evaluations (denoted with "?").

\begin{table}[ht] \begin{center}
\caption{Association of users with requirements $req_1 .. req_5$. \vspace{0.1cm}} 
{\begin{tabular}{|c|c|c|c|c|}
\hline 
relevance & user$_{1}$ &  user$_{2}$ & user$_{3}$ &  user$_{4}$\tabularnewline
\hline
\hline 
req$_{1}$ & ? & ? & 5 & ?\tabularnewline
\hline 
req$_{2}$ & 10 & ? & 1 & ?\tabularnewline
\hline 
req$_{3}$ & ? & 6 & ? & 2\tabularnewline
\hline 
req$_{4}$ & ? & ? & 3 & ?\tabularnewline
\hline 
req$_{5}$ & ? & ? & ? & 5\tabularnewline
\hline 
\end{tabular}} \end{center}
\label{tab:useritemratingtable} 
\end{table}

Based on the information included in Table \ref{tab:useritemratingtable}, we can perform so-called dimensionality reduction and describe the relationship between users and requirements in terms of two low-dimensional matrices $U$ and $R$ where the former describes the relationship between users and abstract dimensions (hidden features) (see Table \ref{tab:userdimensiontable}) and the latter the relationship between items and abstract dimensions (see Table \ref{tab:itemdimensiontable}).

\begin{table}[ht] \begin{center}
\caption{ User $\times$ interest dimension ($d_1 .. d_3$) affinity matrix $U$. \vspace{0.1cm}} 
{\begin{tabular}{|c|c|c|c|c|}
\hline 
 & user$_{1}$ &  user$_{2}$ & user$_{3}$ &  user$_{4}$\tabularnewline
\hline
\hline 
d$_{1}$ & 3,652807135 & 1,251029912 & 0,148850849 & 1,870385191\tabularnewline
\hline 
d$_{2}$ & 2,406538532 & 1,830201936 & 1,766613942 & 0\tabularnewline
\hline 
d$_{3}$ & 0,053547355 & 0,176813763 & 1,86544824 & 0,002507298\tabularnewline
\hline 
\end{tabular}} \end{center}
\label{tab:userdimensiontable}
\end{table}

The table entries can be learned on the basis of a matrix factorization approach that is based on non-linear optimization. The optimization goal is to find values for the low-dimensional tables, which help to predict the missing table entries as good as possible. For a detailed discussion of matrix factorization techniques we refer to \cite{Koren2009}.

Similar to the description of the relationship between users and hidden features, we can describe the relationship between requirements and hidden features. The higher the value, the higher the corresponding affinity between users (requirements) and the corresponding hidden features. We want to emphasize that in the matrix factorization context features are \emph{hidden}, i.e., it is not clear if and which hidden feature corresponds to a specific evaluation dimension (as discussed in the context of utility-based prioritization).
			
\begin{table}[ht] \begin{center}
\caption{ Requirement $\times$ interest dimension ($d_1 .. d_3$) affinity matrix $R$. \vspace{0.1cm}}
{\begin{tabular}{|c|c|c|c|c|}
\hline 
 & d$_{1}$ &  d$_{2}$ & d$_{3}$ \tabularnewline
\hline
\hline 
req$_{1}$ & 0,318390415 & 0,359262854 & 2,305033956 \tabularnewline
\hline 
req$_{2}$ & 2,527786478 & 0,3177104899 & 0,035500999 \tabularnewline
\hline 
req$_{3}$ & 1,072394897 & 2,524779729 & 0,126403403 \tabularnewline
\hline 
req$_{4}$ & 0,167185814 & 1,181561695 & 0,467019398 \tabularnewline
\hline 
req$_{5}$ & 2,665424355 & 0,109392275 & 0,008631143 \tabularnewline
\hline 
\end{tabular}} \end{center}
\label{tab:itemdimensiontable}
\end{table}

The two low-dimensional matrices $U$ and $R$ can now be used to  calculate a prediction for an unspecified user $\times$ requirement pair denoted with "?" (see Table \ref{tab:useritemratingtable}). By applying matrix multiplication, we can, for example, determine a prediction of the evaluation of requirement $req_1$ by $user_1$. The corresponding table entry results from the expression $0,318390415 \times 3,652807135 + 0,359262854 \times 2,406538532 + 2,305033956 \times 0,053547355$ which is $2,151027154$. Expecting predictions on a scale $0..10$, the prediction for the evaluation of requirement $req_1$ by $user_1$ appears to be rather low.

\section{Minimum Viable Products} \label{sec:mvp}
Minimum viable products (MVPs) represent products (in our case software components) that include a minimum set of requirements applicable and of value for a customer. In the context of software development, MVP development is extremely important especially for start-up companies since resources are often extremely limited and there is only one chance to develop the right product for the customer community. Consequently, prioritization support is extremely important in such scenarios. MVP development is related to DevOps software processes which are characterized by extensive automation and continuous updates \cite{Lwakatare2019}. Such processes support a more in-depth customer integration into feedback and prioritization and -- as a consequence -- help to increase the quality of prioritization due to deeper insights into the progress of the project.

Prioritizations for \emph{minimum viable products} typically have to deal with high-level requirements, which do not describe specific functionalities but rather generic features of the software. For these features, it should be made clear which are the most relevant ones that can realistically be implemented. We can consider the  task of selecting a subset of requirements to be included in a minimal viable product as a utility-based prioritization task where requirement \emph{utilities} and \emph{time estimates} are used as basic inputs in a follow-up process that focuses on optimizing the selection of a bundle of most relevant features (requirements). Thus, MVP-oriented prioritization supports a kind of triage process \cite{20_Davis2003} where the most important and feasible requirements are implemented first.

Formula \ref{eq:timeconstraint} restricts the available time resources, i.e., how much time is available to implement the new MVP features. In typical start-up scenarios, this would reflect a situation where, for example, four persons together can spend around one month to implement market-relevant features into an MVP. To make good use of the available time, resource planning can be used to calculate an optimal subset of requirements to be included ($included(req_i)$) in the MVP. An example of how to take into account  time restrictions is shown in Formula \ref{eq:timeconstraint}.

\begin{equation} \label{eq:timeconstraint}
    time(req_1) \times included(req_1) + .. + time(req_n) \times included(rec_n) \leq maxtime
\end{equation}

The overall optimization objective of this resource planning task is expressed with Formula \ref{eq:optimizeresources}. The utility of the selected requirements (requirements, which should be part of the MVP) should be maximized while taking into account additional restrictions (see Formula \ref{eq:timeconstraint}).

\begin{equation} \label{eq:optimizeresources}
    max \leftarrow utility(req_1) \times included(req_1) + .. + utility(req_n) \times included(rec_n)
\end{equation}

\begin{table}[ht] \begin{center}
\caption{ Selecting the most relevant requirements under given time conditions resulting in a maximum utility of $10.31$ = utility($req_2$)+utility($req_5$). \vspace{0.1cm}}
{\begin{tabular}{|c|c|c|c|c|c|}
\hline 
requirement $req_i$ & req$_{1}$ &  req$_{2}$ & req$_{3}$ &  req$_{4}$ &  req$_{5}$\tabularnewline
\hline
\hline 
utility($req_i$) & 4.63 & \bf5.75 & 4.06 & 2.94 & \bf4.56\tabularnewline
\hline 
time($req_i$) & 3 & 4 & 4 & 3 & 5\tabularnewline
\hline 
selected & 0 & \bf1 & 0 & 0  &  \bf1\tabularnewline
\hline 
\end{tabular}} \end{center}
\label{tab:optimizereqselection}
\end{table}

\section{Basic Release Planning} \label{sec:brp}

Basic release planning follows a prioritization approach where requirements formulated on a fine-granular level are selected with regard to their relevance of being part of one of the next $n$ releases -- in the case of $n=1$, this scenario is also denoted as \emph{next release problem}. In most of the cases, such scenarios do not need the support of a high-sophisticated release planning solution. Example reasons for choosing a lightweight process are the \emph{unavailability of resource data} required by release planning tools (e.g., data about resources already occupied in projects) and \emph{limited budgets and personnel resources} to purchase and support a heavy-weight release planning software and to integrate this software with resource-related data sources.

Basic release planning focuses on the prioritization of requirements formulated on a fine-granular level. Initially, this process is often performed on the basis of a utility analysis (see Section \ref{sec:earlyre}). On the basis of the results of a utility analysis, stakeholders can propose assignments of requirements to releases. If a company's software process follows a \emph{next release} strategy, i.e., the planning horizon is the next release, the corresponding selection task is to figure out the most relevant requirements for the next release. Basic release planning typically does not take into account constraints regarding available resources -- such constraints are taken into account informally. 

Tools supporting basic release planning can help to repair inconsistencies in the stakeholders' preferences regarding the assignment of requirements to releases. A  scenario in the context of basic release planning is the following (see Table \ref{tab:basicreleaseplanning}). Stakeholders (users) define their individual preferences regarding the assignment of requirements to releases. Since stakeholders can do this remotely and are initially often not allowed to see the preferences of other stakeholders, conflicts regarding defined release assignment preferences can occur \cite{Stettinger2015}.

\begin{table}[ht] \begin{center}
\caption{ Preferences of stakeholders with regard to release assignments. \vspace{0.1cm}}
{\begin{tabular}{|c|c|c|c|c|}
\hline 
 & user$_{1}$ &  user$_{2}$ & user$_{3}$ & user$_{4}$\tabularnewline
\hline
\hline 
req$_{1}$ & 1 & 1 & 2 & 1\tabularnewline
\hline 
req$_{2}$ & 2 & 2 & 3 & 3\tabularnewline
\hline 
req$_{3}$ & 3 & 3 & 3 & 3\tabularnewline
\hline 
req$_{4}$ & 1 & 2 & 2 & 3\tabularnewline
\hline 
req$_{5}$ & 4 & 1 & 1 & 1\tabularnewline
\hline 
\end{tabular}} \end{center}
\label{tab:basicreleaseplanning}
\end{table}

In this context, constraint-based optimization can be applied to minimize the need of preference change per user (see Formula \ref{eq:consistentreleaseassignment}). We assume the existence of variables $ureq_{ij}$ with the domain $1..4$ representing the releases $1..4$, for example, $ureq_{11}=1$ indicates that $user_1$ prefers the assignment of $req_1$ to release 1. Furthermore, we assume the existence of variables $ureq'_{ij}$, which represent the solution space. The constraint $ureqcount_{ij} = abs(ureq_{ij}-ureq'_{ij})$ indicates whether a user preference has to be adapted. Furthermore, we need to count the number of changes needed per user $i$  (see Formula \ref{eq:stakeholdercount}). The number of  preference changes per user $i$ is represented by variable $chn_i$ (see Formula \ref{eq:stakeholdercount}). 

\begin{equation} \label{eq:stakeholdercount}
    chn_i \leftarrow ureqcount_{i1} + .. + ureqcount_{in}
\end{equation}

Furthermore, we want to assure \emph{consensus}, i.e., each requirement $j$ has to be assigned to exactly one release (see Formula \ref{eq:equivalentassignment}).

\begin{equation} \label{eq:equivalentassignment}
    ureq'_{1j} = .. = ureq'_{mj}
\end{equation}

Given this knowledge, we can define an optimization problem with the overall goal to minimize the number of changed release assignments while at the same time being fair, i.e., it should not be the case that (in the worst case) all needed changes are affecting a single stakeholder. This criteria is represented by Formula \ref{eq:consistentreleaseassignment}. The underlying idea is that the pairwise distance between stakeholders in terms of the number of needed stakeholder-specific preference adaptations should be minimized. 

\begin{equation} \label{eq:consistentreleaseassignment}
    min \leftarrow abs(chn_1 - chn_{2}) +  .. + abs(chn_{n-1} - chn_{n})
\end{equation}

Formula \ref{eq:consistentreleaseassignmenttradeoff} represents an alternative optimization function where the expected solution represents a tradeoff between \emph{fairness} among stakeholders in terms of a fair share of individual changes of preferences and \emph{minimality} in terms of the overall number of needed changes.

\begin{equation} \label{eq:consistentreleaseassignmenttradeoff}
    min \leftarrow (abs(chn_1 - chn_{2}) +  .. + abs(chn_{n-1} - chn_{n})) \times (chn_1 + .. + chn_n)
\end{equation}

This kind of knowledge  can be exploited by optimization features of constraint solvers such as \textsc{Choco}.\footnote{choco-solver.org}

\section{Integrated Release Planning} \label{sec:irp}

On top of the concepts of basic release planning, integrated release planning has a strong focus on integrating additional constraints related to the dependency between requirements and  constraints related to the availability of resources, limits of resource consumption, and the assignment of stakeholders to individual tasks. Integrated release planning requires detailed information about the assignment of employees to current projects and their availability. Furthermore, project-specific release plans have to be synchronized since employees can be assigned to multiple projects during the same time period. A special case are distributed project scenarios where a large project is conducted by different independent teams that work on some common features, which have to be taken into account in the release plans of the individual project partners.

Table \ref{tab:detailedreleaseplanning} provides a representative overview of modeling concepts that can be used in the context of release planning. Requirements can be represented as basic components with associated properties represented as finite domain variables. For example, $req_1.rel$ denotes requirement $req_1$ with the associated release $req_1.rel$, which could be represented, for example, by the domain 1..3, i.e., the look-ahead factor for releases would be 3. Another example of a property which can be associated with a requirement $req_i$ is $req_i.dur$, which denotes the time estimate for requirement $req_i$.

\begin{table}[ht] \begin{center}
\caption{ Examples of basic constraints used for defining release planning tasks. In this context, $req_i$ denotes a requirement, $req_i.rel$ denotes the corresponding release, and $req.dur$ denotes the estimated development time for a requirement. \vspace{0.1cm}}
{\begin{tabular}{|c|c|c|c|c|}
\hline 
 Definition &   Description\tabularnewline
\hline
\hline 
$req_i.rel = a$ & $req_i$ is assigned to release $a$\tabularnewline
\hline 
$req_i.rel < req_j.rel$ & $req_i$ must be implemented before  $req_j$\tabularnewline
\hline 
$req_i.rel \leq req_j.rel$ & $req_j$ must not be implemented before  $req_i$\tabularnewline
\hline 
$req_i.rel \neq req_j.rel$ & $req_i$ and $req_j$ must have different releases\tabularnewline
\hline 
$req_i.rel \leq a$ &  implementation of $req_i$ not after release $a$ \tabularnewline
\hline 
$req_i.rel \geq a$  & implementation of $req_i$ not before release $a$\tabularnewline
\hline 
$req_i.rel = n \lor req_j.rel = n$  & $req_i$ or $req_j$ not in release plan\tabularnewline
\hline
$\neg(|req_i.rel - req_j.rel| > k)$  & $req_i$ and $req_j$ must be implemented  timely \tabularnewline
\hline 
$|\{r \in R: r =  rel\}| \leq a$  & not more than $a$ requirements in release $rel$\tabularnewline
\hline
$\Sigma_{r \in R \land r.rel=rel} (r.dur) \leq a$  & not more than $a$  hours bounded to $rel$\tabularnewline
\hline
\end{tabular}} \end{center}
\label{tab:detailedreleaseplanning}
\end{table}

A simple example of the application of the modeling concepts shown in Table \ref{tab:detailedreleaseplanning} is given in Tables \ref{tab:exampleofdependencies}--\ref{tab:exampleofpreferences}. Table \ref{tab:exampleofdependencies} includes dependencies between requirements that are considered correct and have to be taken into account, i.e., the constraints are so-called \emph{hard constraints}. For example $req_1.rel < req_2.rel$ denotes the fact that the implementation of $req_1$ has to be completed before the implementation of $req_2$ can be started. Since these constraints are assumed to be taken into account, they have to be consistent, i.e., at least one solution should exist. Assuming a finite domain of $1..3$ for each individual variable $req_i.rel$, a corresponding consistent variable assignment (solution) is $\{req_{1}.rel = 1, req_{2}.rel = 2, req_{3}.rel = 3, req_{4}.rel = 3, req_{5}.rel = 1\}$. 

\begin{table}[ht] \begin{center}
\caption{ Example requirements and set $D$ of corresponding dependencies. The domain of $req_i.rel$ is assumed to be $1..3$. \vspace{0.1cm}} 
{\begin{tabular}{|c|c|c|c|c|c|}
\hline 
     & req$_{1}.rel$ & req$_{2}.rel$ & req$_{3}.rel$ & req$_{4}.rel$ & req$_{5}.rel$ \tabularnewline
\hline
\hline 
req$_{1}.rel$ & - & $<$ & - & - & -\tabularnewline
\hline 
req$_{2}.rel$ & - & - & $<$ & - & $>$\tabularnewline
\hline 
req$_{3}.rel$ & - & - &-  & - & -\tabularnewline
\hline 
req$_{4}.rel$ & - & - & - & - & $\neq$\tabularnewline
\hline 
req$_{5}.rel$ & - & - & - & - & -\tabularnewline
\hline 
\end{tabular}} \end{center}
\label{tab:exampleofdependencies}
\end{table}

\begin{table}[ht] \begin{center}
\caption{Example set $S$ of (inconsistent) stakeholder preferences. \vspace{0.1cm}}
{\begin{tabular}{|c|c|c|c|c|}
\hline 
 & user$_{1}$ &  user$_{2}$ & user$_{3}$ & user$_{4}$\tabularnewline
\hline
\hline 
req$_{1}.rel$ & $=1$ & $=1$ & $\leq 2$ & $=1$\tabularnewline
\hline 
req$_{2}.rel$ & $\geq 2$ & $\geq 2$ & $\geq 2$ & $\geq 2$\tabularnewline
\hline 
req$_{3}.rel$ & $\leq 2$ & $\geq 2$ & $=3$ & $\leq3$\tabularnewline
\hline 
req$_{4}.rel$ & $\geq 1$ & $\geq 1$ & $\geq 2$ & $\geq 2$\tabularnewline
\hline 
req$_{5}.rel$ & $\geq 2$ & $=1$ & $=1$ & $\leq 2$\tabularnewline
\hline 
\end{tabular}} \end{center}
\label{tab:exampleofpreferences} 
\end{table}

Please note that all constraint types shown in Table \ref{tab:basicreleaseplanning} can be either represented as \emph{hard constraints} or as \emph{soft constraints} -- in the context of our example, the entries of Table \ref{tab:exampleofdependencies} are interpreted as hard constraints, those of Table \ref{tab:exampleofpreferences} as soft constraints, i.e., stakeholder preferences that should be taken into account but could also be ignored in the case that not all stakeholder preferences could be taken into account. On the basis of the (hard) constraints shown in Table \ref{tab:exampleofdependencies}, stakeholders (users) can specify their individual preferences (see Table \ref{tab:exampleofpreferences}). For simplicity, we restrict the constraint type of user preferences to the form $req_i.rel = a$, $req_i.rel < a$, $req_i.rel > a$, $req_i.rel \leq a$, and $req_i.rel \geq a$.

The stakeholder preferences $S$ in Table \ref{tab:exampleofpreferences} are inconsistent. Detailed release planning can be regarded as an interactive process where stakeholders define their preferences and then try to establish consensus with regard to the final release plan. In the example shown in Table \ref{tab:exampleofpreferences}, the stakeholders have defined inconsistent preferences with regard to the requirements $req_3$ and $req_5$. More precisely, there is one set of conflicting preferences with regard to $req_3$ ($\{\{user_1: (\leq 2), user_3: (= 3)\}\}$) and two conflicting preferences with regard to $req_5$ ($\{\{user_1: (\geq 2), user_2: (= 1)\}, \{user_1: (\geq 2), user_3: (= 1)\}\}$). Combinations of preferences that induce an inconsistency are often denoted as conflict set \cite{52_Junker2004,FelfernigSchubertZehentner2012}. Conflict sets can be shown to stakeholders to indicate open issues and to stimulate discussions on how to resolve the existing inconsistencies. In our example, the inconsistent situation could be resolved if stakeholder $user_1$ would agree to change both of his (her) preferences. If we take into account both, the constraints in $D$ and the preferences in $S$, we can detect two singleton conflicts both induced by the preferences of $user_1$ ($\{\{user_1: (\leq 2)\}, \{user_1: (\geq 2)\}\}$).

\section{Stakeholder Recommendation} \label{sec:stakeholderrecommendation}
An issue in different prioritization scenarios is to figure out who should be in charge of validating a specific requirement since (s)he has the expertise needed. The quality of stakeholder/requirement assignment can have enormous impacts on the quality of a prioritization since sub-optimal evaluations can lead to sub-optimal prioritizations. Specifically, missing expertise can lead to situations where, for example, requirements of high relevance are evaluated as less relevant and -- as a consequence -- are not considered as a potential candidate for early releases. A major issue is to identify stakeholders who have the expertise and thus can provide reasonable evaluations of requirements. As sketched in Formula \ref{eq:stakeholderreqsimilarity}, expertise estimation can be implemented on the basis of the similarity between requirements already evaluated by a stakeholder and a set of new requirements. 

Stakeholder expertise can be modeled in various ways. In the following, we provide a basic example of how to exploit the concepts of content-based recommendation \cite{77_Pazzani1997} to propose reasonable assignments of stakeholders to requirements. Table \ref{tab:requirementsandkeywords} contains a set of new requirements with a corresponding set of keywords, which have been extracted from the requirement description. For these requirements, we would like to figure out automatically, which stakeholder would be the best one to work on this requirement, for example, to evaluate the requirement. Furthermore, Table \ref{tab:stakeholdersandexpertise} shows a list of stakeholders (users) and a corresponding list of keywords extracted from requirements descriptions the stakeholder worked on in the past. In order to estimate which stakeholder should work on which requirement, we can apply the concepts of content-based recommendation \cite{77_Pazzani1997}. We can calculate the similarity between the keywords describing a stakeholder (see Table \ref{tab:stakeholdersandexpertise}) and the keywords describing a requirement (see Table \ref{tab:requirementsandkeywords}). This can be achieved by applying Formula \ref{eq:stakeholderreqsimilarity} which helps to determine the  stakeholder $\times$ requirements similarity. 

\begin{equation}\label{eq:stakeholderreqsimilarity}
sim(user,req) = \frac{2 \times |keywords(user) \cap keywords(req)|}{keywords(user) \cup keywords(req)}
\end{equation}

\begin{table}[ht] \begin{center}
\caption{Requirements and keywords extracted from their descriptions. \vspace{0.1cm}}
{\begin{tabular}{|c|c|c|c|c|}
\hline 
 Requirements & Keywords \tabularnewline
\hline
\hline 
req$_{1}$ &  registration users\tabularnewline
\hline 
req$_{2}$ & basic payment \tabularnewline
\hline 
req$_{3}$ & credit card payment  \tabularnewline
\hline 
req$_{4}$ & optimize user portfolio\tabularnewline
\hline 
req$_{5}$ & optimize database \tabularnewline
\hline 
\end{tabular}} \end{center}
\label{tab:requirementsandkeywords} 
\end{table}

\begin{table}[ht] \begin{center}
\caption{ Stakeholders and keywords of requirements they have validated. \vspace{0.1cm}}
{\begin{tabular}{|c|c|c|c|c|}
\hline 
 Stakeholders & Keywords \tabularnewline
\hline
\hline 
user$_{1}$ &  registration feature database connection \tabularnewline
\hline 
user$_{2}$ & payment process \tabularnewline
\hline 
user$_{3}$ & credit card interfaces  \tabularnewline
\hline 
user$_{4}$ & credit card portfolio optimize\tabularnewline
\hline 
\end{tabular}} \end{center}
\label{tab:stakeholdersandexpertise}
\end{table}

The result of this similarity evaluation is summarized in Table \ref{tab:contentbasedsimilarities}. For $req_1$ and $req_5$, users with an average similarity have been identified as candidates for validating the requirements. A user with a stronger similarity could be found for $req_2$. Finally, there is a strong similarity between requirements $req_3$, $req_4$, and $user_4$. Overall, $user_4$ seems to have a high coverage with regard to the potential requirements assignments. Finally, $user_3$ has the highest expertise with regard to a single requirement ($req_3$).

\begin{table}[ht] \begin{center}
\caption{Content-based similarity between stakeholders and requirements. \vspace{0.1cm}}
{\begin{tabular}{|c|c|c|c|c|}
\hline 
 & user$_{1}$ &  user$_{2}$ & user$_{3}$ & user$_{4}$\tabularnewline
\hline
\hline 
req$_{1}$ & \bf0.4 & 0 & 0 & 0\tabularnewline
\hline 
req$_{2}$ & 0 & \bf0.66 & 0 & 0\tabularnewline
\hline 
req$_{3}$ & 0 & 0.5 & \bf1.0 & \bf0.8\tabularnewline
\hline 
req$_{4}$ & 0 & 0 & 0 & \bf0.8\tabularnewline
\hline 
req$_{5}$ & \bf0.4 & 0 & 0 & \bf0.4\tabularnewline
\hline 
\end{tabular}} \end{center}
\label{tab:contentbasedsimilarities}
\end{table}

\section{Research Issues} \label{sec:researchissues}

\emph{Derivation of Preferences from Social Networks}. In the discussed prioritization scenarios, preference elicitation is still a manual process. Especially in contexts where companies have established a social network representing their user community, network contents, for example, in the form of tweets can be exploited to infer new requirements and preferences with regard to existing and future software features \cite{Williams2017}. The automated integration of community preferences into requirements prioritization is still an open issue and extremely relevant for making related decision processes more community-oriented and efficient. Beyond automated preference integration, quality assurance for preferences is an extremely important issue. \cite{nayebi2018asymmetric} show how a consequence-based evaluation of different choice alternatives can help to improve the overall quality of release planning decisions.

\emph{Avoidance of Decision Biases}. Decision biases are related to shortcuts in decision making that can lead to sub-optimal decisions \cite{felfernigbolzano2014,FelfernigfMandlSchubert2010}. Being aware of such biases helps to improve the overall quality of decisions processes. An example of such a bias is \emph{anchoring} where the item evaluations of one user that are already visible to other users who haven't evaluated the item up to now, can have an impact on the evaluation behavior of other users \cite{Stettinger2015}. For an overview of decision biases in recommender systems we refer to \cite{felfernigbolzano2014}. Many of the existing biases reported in the psychological literature have not been evaluated up to now. This can be regarded as a major topic for future research.

\emph{Transparency of Decisions}. In order to increase trust, decisions have to be made transparent. Transparency can be achieved on the basis of explanations, which help to understand the reasons for a recommended decision \cite{Felfernig2018}. An important role of transparency is also related to the task of avoiding manipulations in decision making \cite{Tran2019}. An example thereof is a situation where a user tries to adapt his/her rating in order to push his/her preferred alternatives (\emph{push attack}). As discussed in Trang et al. \cite{Tran2019}, a very effective way of avoiding manipulations is to make the rating behavior of individual users more transparent, i.e., making their rating behavior visible to other users. A research issue in this context is to analyze in detail which degree of transparency of rating behavior best helps to counteract manipulations and which visualizations should be used to explain the current status of a decision process.

\emph{Prioritization and Decision Making in Open Source Environments}. Open source development often takes place in the context of single user (contributor) decision making, i.e., contributors can individually and independently  decide which requirement to implement next. Often, many new requirements are potential candidates and the analysis of these candidates is time-consuming. In this context, prioritization can help to automatically rank new requirements in a contributor-specific fashion and thus to significantly reduce related analysis efforts. An approach to support such prioritization scenarios in the \textsc{Eclipse} open source environment is reported in \cite{Felfernig2018RE}. A research challenge in this context is to develop decision support approaches that do not only determine recommendations for individuals but also to figure out which prioritization helps to make the open source community as a whole more productive.

\section{Conclusions} \label{sec:conclusions}
In this paper, we provide an overview of prioritization scenarios that can be differentiated with regard to the degree of underlying requirement granularity and whether constraints are used to describe a prioritization task. These scenarios range from \emph{early requirements engineering} (utility analysis of high-level requirements), \emph{minimum viable product} (selection of features to be contained in a first version of a product), \emph{basic release planning} (initial prioritization of requirements), to \emph{integrated release planning} (detailed prioritization of requirements with regard to a predefined set of releases). To better show the application of related decision support techniques, we introduce a couple of prioritization examples. This paper is concluded with an outline of open issues for future research.

\section*{Acknowledgment}
The work presented in this paper has been conducted within the scope of the Horizon2020 \textsc{OpenReq} Project (funded by the European Union).


\bibliographystyle{plain}   
\bibliography{ws-book-prioritization}

\begin{thebibliography}{10}

\bibitem{Achimugu2014}
P.~Achimugu, A.~Selamat, R.~Ibrahim, and M.~Mahrin.
\newblock A systematic literature review of software requirements
  prioritization research.
\newblock {\em Information and Software Technology}, 56(6):568--585, 2014.

\bibitem{Adomavicius2010}
G.~Adomavicius, N.~Manouselis, and Y.~Kwon.
\newblock {\em Recommender Systems Handbook}, chapter Multi-Criteria
  Recommender Systems, pages 769--803.
\newblock Springer, 1st edition, 2010.

\bibitem{Alenezi2013}
M.~Alenezi and S.~Banitaan.
\newblock Bug reports prioritization: Which features and classifier to use?
\newblock In {\em 12th International Conference on Machine Learning and
  Applications}, pages 112--116, 2013.

\bibitem{Ameller2017}
D.~Ameller, C.~Farre, X.~Franch, D.~Valerio, and A.~Cassarino.
\newblock Towards continuous software release planning.
\newblock In {\em 24th IEEE International Conference on Software Analysis,
  Evoluation and Reengineering (SANER)}, pages 402--406, 2017.

\bibitem{Atas2018liquid}
M.~Atas, T.~Tran, R.~Samer, A.~Felfernig, and M.~Stettinger.
\newblock Liquid democracy in group-based configuration.
\newblock In {\em Workshop on Configuration}, pages 93--98, Graz, Austria,
  2018. CEUR.

\bibitem{20_Davis2003}
A.~Davis.
\newblock The art of requirements triage.
\newblock {\em IEEE Computer}, 36(3):42--49, 2003.

\bibitem{Dyer2005}
J.~Dyer.
\newblock Multi attribute utility theory.
\newblock {\em International Series in Operations Research and Management
  Science}, 78:265--292, 1997.

\bibitem{felfernigbolzano2014}
A.~Felfernig.
\newblock Biases in decision making.
\newblock In {\em Proceedings of the International Workshop on Decision Making
  and Recommender Systems 2014}, volume 1278, pages 32--34, Bolzano, Italy,
  2014. CEUR Proceedings.

\bibitem{Felfernig2018}
A.~Felfernig, L.~Boratto, M.~Stettinger, and M.~Tkalcic.
\newblock {\em Group Recommender Systems -- An Introduction}.
\newblock Springer, 2018.

\bibitem{felfernigburke08}
A.~Felfernig and R.~Burke.
\newblock Constraint-based recommender systems: Technologies and research
  issues.
\newblock In {\em ACM International Conference on Electronic Commerce
  (ICEC08)}, pages 17--26, Innsbruck, Austria, 2008.

\bibitem{FelfernigfMandlSchubert2010}
A.~Felfernig, W.~Maalej, M.~Mandl, M.~Schubert, and F.~Ricci.
\newblock Recommendation and decision technologies for requirements
  engineering.
\newblock In {\em ICSE 2010 Workshop on Recommender Systems in Software
  Engineering}, pages 1--5, Cape Town, South Africa, 2010.

\bibitem{FelfernigSchubertZehentner2012}
A.~Felfernig, M.~Schubert, and C.~Zehentner.
\newblock {An Efficient Diagnosis Algorithm for Inconsistent Constraint Sets}.
\newblock {\em Artificial Intelligence for Engineering Design, Analysis, and
  Manufacturing (AIEDAM)}, 26(1):175--184, 2012.

\bibitem{Felfernig2018RE}
A.~Felfernig, M.~Stettinger, M.~Atas, R.~Samer, J.~Nerlich, S.~Scholz,
  J.~Tiihonen, and M.~Raatikainen.
\newblock Towards utility-based prioritization of requirements in open source
  environments.
\newblock In {\em 26th IEEE Conference on Requirements Engineering}, pages
  406--411, Banff, Canada, 2018. IEEE.

\bibitem{Firesmith2004}
D.~Firesmith.
\newblock {Prioritizing Requirements}.
\newblock {\em Journal of Object Technology}, 3(8):35--47, 2004.

\bibitem{Huang2011}
S.~Huang.
\newblock {Designing utility-based recommender systems for e-commerce:
  Evaluation of preference elicitation methods}.
\newblock {\em Electronic Commerce Research and Applications}, 10(4):398--407,
  2011.

\bibitem{52_Junker2004}
U.~Junker.
\newblock {\textsc{QuickXplain}: Preferred Explanations and Relaxations for
  Over-Constrained Problems}.
\newblock In {\em 19th National Conference on AI (AAAI04)}, pages 167--172, San
  Jose, CA, 2004.

\bibitem{karim2014bi}
Muhammad~Rezaul Karim and Guenther Ruhe.
\newblock Bi-objective genetic search for release planning in support of
  themes.
\newblock In {\em Proceedings Symposium on Search Based Software Engineering},
  pages 123--137. Springer, 2014.

\bibitem{Karlsson1997}
J.~Karlsson and K.~Ryan.
\newblock {A Cost-Value Approach for Prioritizing Requirements}.
\newblock {\em IEEE Software}, 14(5):67--74, 1997.

\bibitem{Kifetew2017}
F.~Kifetew, A.~Susi, D.~Mutante, A.~Perini, A.~Siena, and P.~Busetta.
\newblock Towards multi-decision-maker requirements prioritisation via
  multi-objective optimisation.
\newblock In {\em Forum and Doctoral Consortium Papers Presented at the 29th
  International Conference on Advanced Information Systems Engineering
  (CAiSE'17)}, pages 137--144, Essen, Germany, 2017.

\bibitem{Koren2009}
Y.~Koren, R.~Bell, and C.~Volinsky.
\newblock Matrix factorization techniques for recommender systems.
\newblock {\em IEEE Computer}, 42(8):30--37, 2009.

\bibitem{Lehtola2004}
L.~Lehtola, M.~Kauppinen, and S.~Kujala.
\newblock Requirements prioritization challenges in practice.
\newblock In {\em 5th International Conference On Product Focused Software
  Process Improvement (PROFES)}, pages 497--508, Kansai Science City, Japan,
  2004.

\bibitem{Lwakatare2019}
L.~Lwakatare, T.~Kilamo, T.~Karvonen, T.~Sauvola, V.~Heikkiläc, J.~Itkonen,
  P.~Kuvaja, T.~Mikkonen, M.~Oivo, and C.~Lassenius.
\newblock {DevOps in practice: A multiple case study of five companies}.
\newblock {\em Information and Software Technology}, 114:217--230, 2019.

\bibitem{Mobasher2011}
B.~Mobasher and J.~Cleland-Huang.
\newblock {Recommender Systems in Requirements Engineering}.
\newblock {\em AI Magazine}, 32(3):81--89, 2011.

\bibitem{nayebi2015analytical}
M~Nayebi and G~Ruhe.
\newblock Analytical product release planning.
\newblock In {\em The Art and Science of Analyzing Software Data}, pages
  550--580. Morgan Kaufmann, 2015.

\bibitem{nayebi2018asymmetric}
Maleknaz Nayebi and Guenther Ruhe.
\newblock Asymmetric release planning: Compromising satisfaction against
  dissatisfaction.
\newblock {\em IEEE Transactions on Software Engineering}, 45(9):839--857,
  2018.

\bibitem{Felfernigetal2014}
G.~Ninaus, A.~Felfernig, M.~Stettinger, S.~Reiterer, G.~Leitner, L.~Weninger,
  and W.~Schanil.
\newblock Intellireq: Intelligent techniques for software requirements
  engineering.
\newblock In {\em European Conference on Artificial Intelligence, Prestigious
  Applications of Intelligent Systems (PAIS)}, pages 1161--1166, 2014.

\bibitem{77_Pazzani1997}
M.~Pazzani and D.~Billsus.
\newblock Learning and revising user profiles: The identification of
  interesting web sites.
\newblock {\em Machine Learning}, 27:313--331, 1997.

\bibitem{Perini2009}
A.~Perini, F.~Ricca, and A.~Susi.
\newblock {Tool-supported requirements prioritization: Comparing the AHP and
  CBRank methods}.
\newblock {\em Information and Software Technology}, 51(6):1021--1032, 2009.

\bibitem{88_Ruhe2005}
G.~Ruhe and M.~Saliu.
\newblock The art and science of software release planning.
\newblock {\em IEEE Software}, 22(6):47--53, 2005.

\bibitem{ruhe2002software}
G{\"u}nther Ruhe.
\newblock Software engineering decision support--a new paradigm for learning
  software organizations.
\newblock In {\em International Workshop on Learning Software Organizations},
  pages 104--113. Springer, 2002.

\bibitem{ruhe2010product}
G{\"u}nther Ruhe.
\newblock {\em Product release planning: methods, tools and applications}.
\newblock CRC Press, 2010.

\bibitem{Stettinger2015}
M.~Stettinger, A.~Felfernig, G.~Leitner, and S.~Reiterer.
\newblock Counteracting anchoring effects in group decision making.
\newblock In {\em 23rd Conference on User Modeling, Adaptation, and
  Personalization (UMAP'15)}, volume 9146 of {\em LNCS}, pages 118--130,
  Dublin, Ireland, 2015. Springer.

\bibitem{Tran2019}
T.~Tran, A.~Felfernig, V.~Le, M.~Atas, M.~Stettinger, and R.~Samer.
\newblock User interfaces for counteracting decision manipulation in group
  recommender systems.
\newblock In {\em 27th ACM Conference on User Modeling, Adaptation and
  Personalization (UMAP)}, pages 93--98, Larnaca, Cyprus, 2019.

\bibitem{Tsang1993}
E.~Tsang.
\newblock {\em Foundations of Constraint Satisfaction}.
\newblock Academic Press, London, 1993.

\bibitem{Wiegers2003}
K.~Wiegers.
\newblock {\em Software Requirements}.
\newblock Microsoft Press, 2003.

\bibitem{Williams2017}
G.~Williams and A.~Mahmoud.
\newblock Mining twitter feeds for software user requirements.
\newblock In {\em 25th International Requirements Engineering Conference (RE)},
  pages 1--10, Lisbon, Portugal, 2017. IEEE.

\bibitem{Xuan2012}
J.~Xuan, H.~Jiang, Z.~Ren, and W.~Zou.
\newblock Developer prioritization in bug repositories.
\newblock In {\em 34th International Conference on Software Engineering
  (ICSE)}, pages 25--35, Z\"urich, Switzerland, 2012.

\end{thebibliography}

\end{document}